%% file: main.tex
\documentclass[pamm,a4paper,fleqn]{w-art}
\usepackage{times,cite,w-thm,amsmath,amssymb,amsfonts}
\usepackage[T1]{fontenc}
\usepackage[utf8]{inputenc}
\usepackage{tensor}
\usepackage{hyperref}
\usepackage[english]{babel}
\usepackage{graphicx}
\input{symbols.tex}

\makeatletter
\renewcommand{\ps@firstpage}{\ps@empty}
\@abstractcopyrightfalse
\makeatother
\begin{document}

\TitleLanguage[EN]
\title[The short title]{Surface Phase-Field-Crystal-Helfrich model for out-of-plane deformations in thin crystalline sheets with lattice mismatch}

\author{\firstname{Emma} \lastname{Radice}\inst{1}%
} 
\address[\inst{1}]{\CountryCode[DE]Institute of Scientific Computing, TU Dresden, Dresden, Germany 
}
\author{\firstname{Ingo} \lastname{Nitschke}\inst{1}%
     }
\author{\firstname{Marco} \lastname{Salvalaglio}\inst{1,2}%
     }
\address[\inst{2}]{\CountryCode[DE]Dresden Center for Computational Materials Science, TU Dresden, Germany
}
\author{\firstname{Axel} \lastname{Voigt}\inst{1,3,4}%
     }
\address[\inst{3}]{\CountryCode[DE]Cluster of Excellence - Physics of Live, TU Dresden, Germany
}
\address[\inst{4}]{\CountryCode[DE]Center for Systems Biology Dresden (CSBD), Germany
}

\AbstractLanguage[EN]
\begin{abstract}
Thin, flexible crystalline sheets exhibit unique elastic properties due to their ability to undergo out-of-plane deformations. Understanding this behavior requires a description that couples in-plane elasticity, out-of-plane deformation, and their coupling, taking the crystalline structure and its defects into account. 
We develop a multiscale description for these systems by extending the surface Phase-Field-Crystal-Helfrich model. The extension permits a spatially varying equilibrium lattice spacing, enabling the representation of localized lattice eigenstrain to mimic lattice mismatch in heterostructures.
We validate the extended model against analytical predictions from classical Föppl–von Kármán equations for uniaxial compression and from Eshelby’s inclusion problem. Using this validated framework, we then show how locally induced compressive stresses drive out-of-plane deformation in the sheets.
\end{abstract}
\maketitle                   

\section{Introduction}
The ability of two-dimensional (2D) materials to undergo out-of-plane deformation gives rise to a wide variety of morphologies and functionalities. This capability has enabled numerous applications, ranging from flexible electronics and stretchable devices to tunable optical and mechanical systems \cite{2DHeterostructures}. Understanding the coupling between in-plane elasticity and out-of-plane deformation is therefore essential for controlling and designing the behavior of crystalline 2D materials. 

Successful descriptions which address the coupling of in-plane stresses and out-of-plane deformations have been proposed using atomistic description, e.g. \cite{Jeong2008,Ariza2010,Cui2020,Torkaman-Asadi2022}, continuum modeling, e.g. \cite{Seung1988,Witten2007,Guinea2008,Roychowdhury2018}, and multiscale approaches, e.g. \cite{aland2012buckling,Zhang2014a,praetorius2018active,PhysRevMaterials.5.034004,BENOITMARECHAL2024105114}. In this work, we follow \cite{PhysRevMaterials.5.034004,BENOITMARECHAL2024105114} and model both the in-plane elastic response and the out-of-plane deformation of crystalline 2D materials retaining the atomistic detail of the crystal. To this end, we employ a surface Phase-Field-Crystal (sPFC) model \cite{Backofen2010,Backofen2011,Kohler2016} coupled with a macroscopic description based on the Helfrich energy to account for bending effects \cite{Helfrich1973}. We extend this model by allowing for spatially varying equilibrium lattice spacing \cite{Punke_2022,SALVALAGLIO2022100067}, enabling the representation of localized lattice eigenstrain to mimic lattice mismatch in heterostructures. We showcase the method by studying the interplay between localized lattice misfit and the presence of defects in heterogeneous crystalline 2D materials. Localized misfit introduces internal stresses that can drive out-of-plane deformation, while defects such as dislocations mediate plastic deformation and contribute additional strain within the crystal, potentially serving as a relaxation mechanism. The competition between these effects determines whether the system relaxes predominantly through defect formation or through out-of-plane deformation. Understanding this interplay is key to predicting and controlling the morphology and mechanical response of 2D crystalline sheets.

Here we formulate the sPFC-Helfrich model using a height observer, discuss model extensions with respect to previous approaches considered in \cite{PhysRevMaterials.5.034004,BENOITMARECHAL2024105114}, and test the results against classical Föppl–von Kármán (FvK) equations under uniaxial compression and with respect to analytical solutions for Eshelby's inclusion problem. To demonstrate the impact of coupling in-plane stresses and out-of-plane deformations, we again consider Eshelby's inclusion problem, but now vary the bending rigidity and incorporate defects by rotating the inclusion. A detailed derivation of the model, a discussion of its analytical properties, and the impact on more general heterostructures will be discussed elsewhere.

\section{Methods}

The Phase-Field-Crystal (PFC) model provides a minimal description of lattice structures evolving over diffusive timescales \cite{Elder2002,Elder2004,Elder2007,Teeffelen2009}. This is achieved by defining an energy functional ${\mathcal{F}}[\psi]$ with $\psi \equiv \psi(\mathbf{x}, t)$ periodic fields approximating the atomic density in crystalline arrangements, and the evolution of $\psi$ given by a conserved gradient flow. The free energy can be written as the Hooke's law upon small deformation of $\psi$ \cite{ElderPRE2010,Heinonen2014}, while defects such as dislocations and grain boundaries forming at interfaces between mismatched/tilted crystals naturally emerge as a result of energy minimization \cite{Elder2002}. The model can also be formulated on curved surfaces, leading to the sPFC model \cite{Backofen2010,Backofen2011,Kohler2016}. All properties of the classical PFC model remain, with the exception that distances are measured as geodesic distances. Also, defects can be present in equilibrium configurations for closed surfaces due to topological constraints. 

The situation changes if the surface is allowed to deform. Instead of a simple conserved surface gradient flow for $\psi$, we now require a gradient flow which minimizes the free energy by simultaneously changing $\psi$ and the surface $\mathcal{S} = \mathcal{S}(t)$, defined by a parametrization $\mathbf{X} = \mathbf{X}(t)$, it is defined on. The mathematical foundations for such problems are still limited. In \cite{NitschkeSadikVoigt_IJoAM_2023}, a general situation has been considered for ($L^2,L^2$) - surface gradient flows. It has been shown that, to guarantee energy dissipation, a notion of surface independence must be chosen consistently with the time derivative. The natural choice is the material time derivative $\dot{\psi}$, see, \eg,  \cite{CERMELLI_FRIED_GURTIN_2005,Dziuk_Elliott_2013,NitschkeVoigt_JoGaP_2023} for proper definitions on evolving surfaces, and the material gauge of surface independence $\eth_{\Wb}\psi=0$, see \cite{NitschkeSadikVoigt_IJoAM_2023}. The situation changes for ($L^2, H^{-1}$) - surface gradient flows, e.g., a $L^2$ - gradient flow with respect to the evolution of ${\mathbf{X}}$ and a $H^{-1}$ - surface gradient flow with respect to the evolution of $\psi$. Besides energy dissipation by evolving ${\mathbf{X}}$ and $\psi$ simultaneously, such flows also should ensure conservation of $\psi$, e.g., $\frac{d}{dt} \int_\mathcal{S} \psi d\mathcal{S} = 0$. At least if local expansion or contraction of the surface is allowed, the natural choice of the material time derivative and the material gauge of surface independence reaches its limits. In \cite{nitschke2026scalar}, the scalar Truesdell time derivative $\mathring{\psi} = \dot{\psi} + \psi \DivC \mathbf{V}$, where $\Div_C$ is the componentwise trace-divergence and $\mathbf{V}$ is the material velocity, and the Truesdell gauge of surface independence 
$\eth_{\Wb}\psi = -\psi\DivC\Wb$, with $\Wb\in\tangentR$, 
are introduced to deal with this situation. A comprehensive overview of the used notation can be found in \cite{Nitschke_2025} and in a condensed form in \cite{nitschke2026scalar}. Together, the definitions allow for constructing ($L^2, H^{-1}$) - surface gradient flows that guarantee energy dissipation and conservation properties. We here follow this general approach for the free energy $\mathcal{F} = \mathcal{F_L} + \mathcal{F_P} + \mathcal{F}_b$ with
\begin{align}
&\mathcal{F_L} = \frac{A}{2} \int_S \left( \psi \mathcal{L}_S\psi\right)d\mathcal{S} \\
&\mathcal{F_P}=\int_S \left( \frac{B}{2} \psi^2+ \frac{C}{3} \psi^3+\frac{D}{4} \psi^4 \right) d\mathcal{S}\\
&\mathcal{F}_b= \frac{\beta}{2}\int _S \mathcal{H}^2 d\mathcal{S}
\end{align}
with $A, B, C, D$ parameters of the sPFC energy, characterizing the phase space and material properties together with the global average density $\bar{\psi} = \int_{\mathcal{S}} \psi d\mathcal{S}$, the operator $\mathcal{L}_\mathcal{S}= (q^2+\Delta_\mathcal{S} )^2$, with $\Delta_\mathcal{S}$ the Laplace-Beltrami operator, approximates the two-point correlation function and encodes the crystal symmetry, which is here chosen to be triangular with $q = q(\mathbf{x})$ modeling a spatially dependent equilibrium lattice parameter \cite{Punke_2022,SALVALAGLIO2022100067}. We thus obtain $\psi\mathcal{L}_S\psi=q^4\psi^2 + q^2\psi\Delta_S\psi+\psi\Delta_S(q^2\psi)+\psi\Delta^2_S\psi$. The last energy contribution is a macroscopic Helfrich energy with $\beta$ the bending rigidity and $\mathcal{H}$ the mean curvature of $\mathcal{S}$. The ($L^2, H^{-1}$) - surface gradient flow reads 
\begin{align}
    M_{\mathbf{X}} \mathbf{V} &= - \mathcal{D}_{\mathbf{X}} {\mathcal{F}} \label{SGF_1}\\
    M_\psi \mathring{\psi} &= \Delta_{\mathcal{S}} \mathcal{D}_{\psi} {\mathcal{F}} \label{SGF_2}
\end{align}
with mobility coefficients $M_\mathbf{X}, M_\psi > 0$ and functional derivatives $\mathcal{D}_{\mathbf{X}} {\mathcal{F}} \in T \mathbb{R}^3 |_\mathcal{S}$ and $\mathcal{D}_{\psi} {\mathcal{F}} \in T^0 \mathcal{S}$. In this form, the surface gradient flow guarantees the conservation property $\frac{d}{dt} \int_\mathcal{S} \psi d\mathcal{S} = 0$ and energy dissipation $\frac{d}{dt} \mathcal{F} \leq 0$. The full set of equations and the proof of the properties will be discussed elsewhere.

Here we restrict the surface evolution to the normal direction, e.g., $M_{\mathbf{X}} \vnor = \normal\DX{\mathcal{F}}$, with $\normal$ the outward pointing normal to $\mathcal{S}$ and $\vnor$ the normal component of $\mathbf{V} = \vnor \normal + \vb$, and consider a height observer given by the parameterization 
\begin{align}
    \para_{\ofrak}:\ 
        (t,x,y)\mapsto \para_{\ofrak}(t,x,y)
        &:= \left[ x, y, h(t,x,y) \right]^T \in\surf\subset\R^3\formComma
\end{align} 
where $h(t,x,y)$ is the scalar height field and $ (x,y)\in\mathcal{U}\subseteq\R^2 $ are local coordinates. For many situations, this is sufficient to capture the key phenomenology and has also been considered in \cite{PhysRevMaterials.5.034004,BENOITMARECHAL2024105114}. We introduce the normal function $\xi$, which measures the signed distance
from the surface along the normal direction. The coupled problem corresponding to \eqref{SGF_1} and \eqref{SGF_2} reads:
\begin{align}
& \partial_t \psi=\left(\partial_t h\right) \left(\frac{\langle\nabla h, \nabla \psi\rangle}{|\boldsymbol{g}|} + \psi \Delta_S h \right)+M_\psi^{-1} \Delta_S\left(\frac{\delta \mathcal{F}}{\delta \psi}\right) \label{eq_heigt1}\\
& \partial_t h=-M_{\mathbf{X}}^{-1} \sqrt{|\boldsymbol{g}|}\left(\frac{\delta \mathcal{F}}{\delta \xi}\right) \label{eq_height2}
\end{align}
with $\boldsymbol{g}$ the metric tensor. The term $\psi \Delta_{\mathcal{S}} h$ in \eqref{eq_heigt1} results from the general derivation in \cite{nitschke2026scalar} and had not been considered in \cite{PhysRevMaterials.5.034004,BENOITMARECHAL2024105114}.  Furthermore, in \cite{PhysRevMaterials.5.034004} only a small-slope approximation is considered, which neglects various geometric nonlinear terms. Further differences can be seen in \eqref{eq_height2}. In \cite{PhysRevMaterials.5.034004,BENOITMARECHAL2024105114} only the elastic parts of $\mathcal{F}$ have been considered to allow for comparison with classical elastic theories. We will follow the same approach and instead of $\mathcal{F}$ only consider $\mathcal{F}_{\mathcal{L}}$ and $\mathcal{F}_b$ in \eqref{eq_height2} in the simulations. Also, the variational derivative $\frac{\delta \mathcal{F}}{\delta \xi}$ differs from \cite{BENOITMARECHAL2024105114} as a result of the considered Truesdell gauge of surface independence that brings in an additional term considering variation with respect to $\psi$. This term will be taken into account after coarse-graining over the atomic unit cell via a smoothing kernel in reciprocal space. Detailed forms of the involved terms formulated in local coordinates $(x,y)\in\mathcal{U}\subseteq\R^2 $ are provided in Appendix \ref{app1}. 

We consider a rectangular domain $(0,L_x) \times (0,L_y)$, periodic boundary conditions, and appropriate initial conditions for $\psi$ and $h$, and solve the system using a Fourier pseudo-spectral method. Nonlinear terms are evaluated in physical space and transformed to Fourier space and back using fast Fourier transforms. This allows for a semi-implicit time-stepping scheme for which an exponential integrator is used. For details and convergence studies on similar approaches, we refer the reader to \cite{BENOITMARECHAL2024105114}.

\section{Numerical results}

\subsection{Validation}

\begin{figure}[h]
    \centering
    \includegraphics[width=1\linewidth]{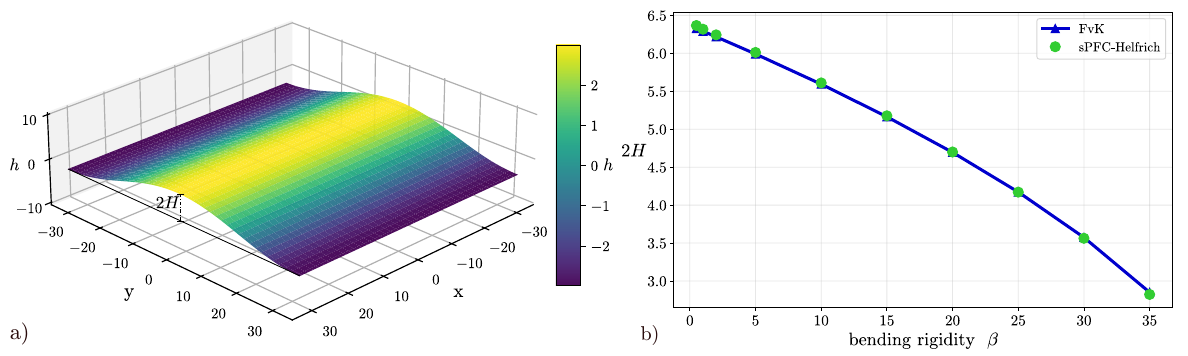}
    \caption{Uniaxial compression. a) Height profile $h$ for bending rigidity $\beta = 5$. b) Height difference $h_{\rm max} - h_{\rm min} = 2 H$ for various $\beta$ together with solution for FvK equation with higher order corrections. The solutions are considered at equilibrium ($t=500,000$). We consider $40$ atoms in the compression direction and use the following parameters: $A = 1$, $B = - 0.2$, $C= 0.5 \sqrt{0.98/3}$, $D = 1$, $M_{\mathbf{X}}=0.001$ and $M_\psi = 1$, following \cite{BENOITMARECHAL2024105114}. The magnitude of the uniaxial strain is $\epsilon = -0.0219$. All lengths are expressed in units of the lattice constant $a = 1/q$.}
   \label{fig:FvK}
\end{figure}

We start by validating the model against simple situations that can also be addressed by more established models, with known or even analytical solutions. We consider the elastic behavior of the sPFC-Helfrich model for a crystalline sheet with a perfect triangular lattice ($q = \text{const}$) subjected to uniaxial compression along $y$. Considering the classical FvK equations for this situation and assuming the first buckling mode $h(y) = H \cos(Qy)$ with $Q = 2\pi/L_y$, an analytical prediction for the amplitude $H$ can be derived by minimizing the elastic free energy with respect to $H$ subject to the boundary condition on the displacement field $u_y(L_y) = \epsilon L_y$, see \cite{Seung1988}. It yields:
\begin{equation}
    (HQ)^2 = -2\!\left(2\epsilon + \frac{2\beta Q^2}{9A\phi^2}\right) = -4(\epsilon - \epsilon_0).
    \label{eq:H_FvK}
\end{equation}
where $\phi = (C + \sqrt{C^2 - 15BD})/(15D)$ is the equilibrium amplitude of the triangular lattice, which relates via $9A\phi^2 = \lambda + 2\mu$ to the elastic constants of an isotropic linear elastic material. We define $\epsilon_0 = -\beta Q^2/(9A\phi^2)$. To compare with \cite{BENOITMARECHAL2024105114}, we consider higher-order correction terms for the critical strain and the surface height gradients leading to:
\begin{equation}
    (HQ)^2 = -4(\epsilon - \epsilon_0) - \frac{7}{6}\!\left(\epsilon^2 + \epsilon_0^2 - \frac{46}{7}\,\epsilon_0\epsilon\right) - \frac{49}{72}\,\epsilon^3 + \frac{19}{2}\,\epsilon^2\epsilon_0 + \frac{245}{24}\,\epsilon_0^2\epsilon - \frac{361}{36}\,\epsilon_0^3 .
    \label{eq:HQ_term}
\end{equation}
Considering as initial condition for $h$ a smooth cosine function in $y$, we solve the full set of equations \eqref{eq_heigt1} and \eqref{eq_height2} until we reach equilibrium for various bending rigidities $\beta$ and measure the amplitude $H$. The solutions and the comparison with the analytic predictions in eq. \eqref{eq:HQ_term} are shown in Figure \ref{fig:FvK}.

\begin{figure}
   \centering
   \includegraphics[width=1\linewidth]{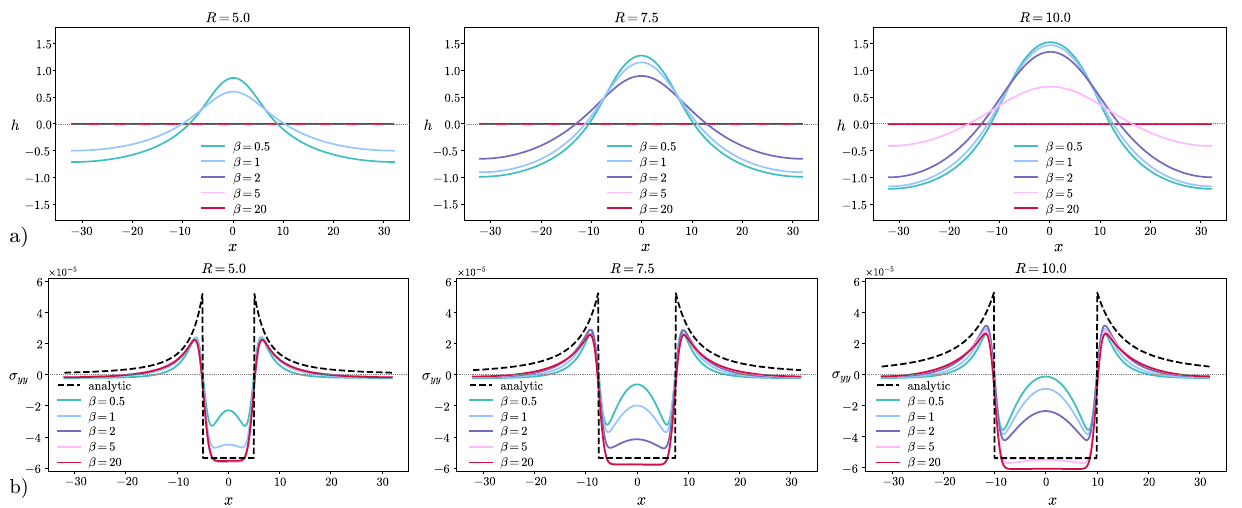}
   \caption{Eshelby inclusion problem. a) Stress field $\sigma_{yy}$ for various radii and bending rigidities. b) Corresponding height profiles $h$. Because upward or downward buckling simply reflects the choice of initial perturbation of the flat sheet, we present only one direction for clarity. The analytical solutions are also reported as dashed lines. Parameters are: $A = 1$, $B = - 0.2$, $C= 0.5 \sqrt{0.98/3}$, $D = 1$, $M_{\mathbf{X}}=0.001$ and $M_\psi = 1$, following \cite{BENOITMARECHAL2024105114}. In addition, we consider the eigenstrain $\epsilon^* = 0.02$ in the inclusion and $\epsilon^* = 0$ in the surrounding matrix, which defines the spatially dependent equilibrium lattice parameter $q(\mathbf{x}) = 1/(1 + \epsilon^*(\mathbf{x})) q_0$. All lengths are expressed in units of the lattice constant $a = 1/q_0$.}
   \label{fig:Eshelby}
\end{figure}

While this essentially validates the treatment of the macroscopic Helfrich energy and the considered coarse-graining, we now turn to the effect of lattice misfit. 
Within the framework of linear continuum elasticity, the stress and strain fields generated by an inclusion with a prescribed eigenstrain are described by Eshelby’s inclusion problem. For an ellipsoidal inclusion, Eshelby showed that the stress (and strain) within the inclusion is uniform, while outside it decays with distance from the interface and can be determined analytically. \cite{Esh110.1098/rspa.1957.0133,esh210.1098/rspa.1959.0173,Esh_mura}. In \cite{SALVALAGLIO2022100067}, this solution has already been used to validate the PFC model with localized eigenstrain. However, this study did not allow for out-of-plane displacements. We reproduce these results considering the full sPFC-Helfrich model for large $\beta$ for spherical inclusions of various radii $R$, see Figure \ref{fig:Eshelby} (plots for $\beta = 20$). The height profile stays constant, while the stress field $\mathbf{\sigma}$, represented by $\sigma_{yy}$, matches the classical Eshelby problem (analytic, dashed line), with the stresses inside the inclusion remaining constant and deviations consistent with previous analyses \cite{SALVALAGLIO2022100067}. The increasing deviations with increasing $R$ are an effect of the boundary conditions. To obtain these results requires computing the stress field from $\psi$, which can be achieved by extending the approach in \cite{salvalaglio2019closing,PhysRevB.103.224107,https://doi.org/10.1002/pamm.202300213} to surfaces. 

\subsection{Results}

Building on the validations discussed above, we next reduce $\beta$ and explore its impact and the size of the inclusion on the amplitude $H$ of the height profile and stress field $\sigma_{yy}$, see Figure \ref{fig:Eshelby}(plots for $\beta =0.5,\ldots, 5$). For moderate $\beta$, the bending of the crystalline sheet provides an additional relaxation mechanism, leading to a partial release of stress within the inclusion. Importantly, different from Eshelby's inclusion problem, or the full sPFC-Helfrich problem with $\beta = 20$, lower bending rigidities introduce a size dependence. Below some threshold, the out-of-plane deformation depends on $R$, and as a result, the stress inside the inclusion also varies with $R$.

\begin{figure}
    \centering
    \includegraphics[width=1\linewidth]{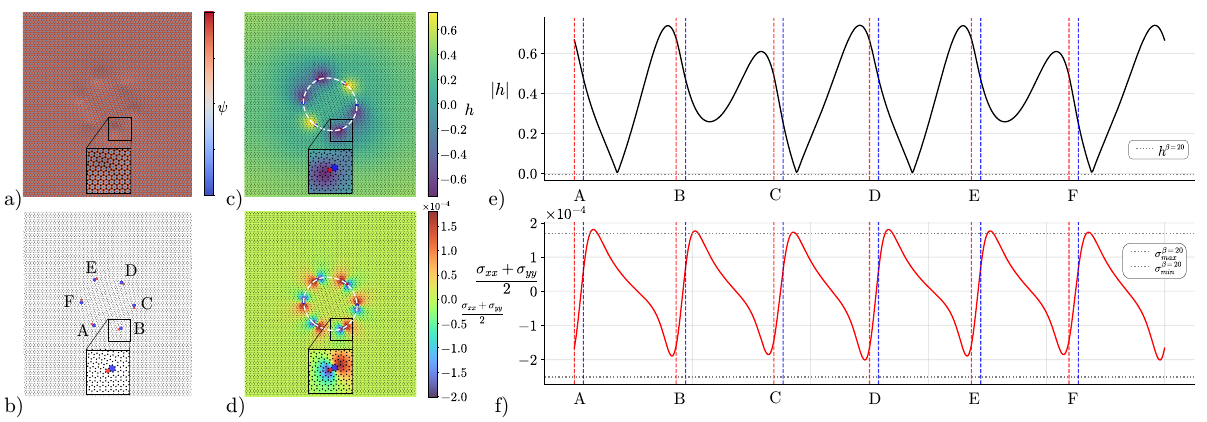}
    \caption{Circular rotated grain. a) Periodic field $\psi$ on the surface $\mathcal{S}$ after relaxation of the initial condition, allowing for defect formation and out-of-plane buckling. b) Extrema of $\psi$ displayed on the surface $\mathcal{S}$ and highlighted dislocations by color coding the regions with a number of neighbors deviating from 6 (blue - 7, red - 5). c) Height profile $h$ shown in $(x,y)$-plane, together with data from b) projected on $(x,y)$-plane. d) Hydrostatic stress $(\sigma_{xx} + \sigma_{yy}) / 2$ together with data from b) projected on $(x,y)$-plane. Insets in a) – d) show zoomed-in views of the region surrounding a single dislocation. The estimated circular interface between the rotated grain and the matrix is shown as white lines in panels c) and d). e) Absolute value of height profile $|h|$ along the interface (white line) in c), with red/blue dashed lines marking the position of the dislocations. f) Hydrostatic stress $(\sigma_{xx} + \sigma_{yy}) / 2$ along the interface (white line) in d), with red/blue dashed lines marking the position of the dislocations. Simulations parameters: $\beta = 0.5$, $R = 10a$ and $\theta=5^\circ$. Corresponding results for $\beta = 20$ are indicated in e) and f) by gray dashed lines. Other parameters are as in Figure \ref{fig:Eshelby}. The grain is placed in the center of the domain.}
    \label{fig:dislocation}
\end{figure}

Next, we consider an inclusion without eigenstrain, but introduce defects by rotating the crystal in the inclusion by an angle $\theta$ with respect to the matrix. This practically introduces a small-angle grain boundary consisting of isolated dislocations. The circular grain shrinks over time, which is well studied for 2D crystals if out-of-plane deformations are prohibited \cite{salvalaglio2019closing,salvalaglio2020coarse}, e.g., in our setting for $\beta = 20$. Here, we are interested in the effect of the dislocations on the height profile for small $\beta$. For $\beta = 0.5$ and the considered rotation leading to 6 dislocations, the surface bulges at the dislocations, see Figure \ref{fig:dislocation} a) - c). In accordance with the results in \cite{BENOITMARECHAL2024105114}, bulges form at the dislocations where the stress is compressive (large values of $|h|$ correspond to small values of $(\sigma_{xx} + \sigma_{yy}) / 2$, see Figure \ref{fig:dislocation} e) and f)). Whether bulging is up or down at every single dislocation depends on the initial perturbation of the height profile. However, the resulting out-of-plane deformation is affected by dislocation-dislocation interaction. The height profile for the considered simulation is shown in Figure \ref{fig:dislocation} e). A different height profile is obtained whether neighboring dislocations have out-of-plane deformations in opposite or the same direction. However, this is found to have a negligible impact on the in-plane hydrostatic stress, see Figure \ref{fig:dislocation} d) and f).

\begin{figure}
    \centering
    \includegraphics[width=1\linewidth]{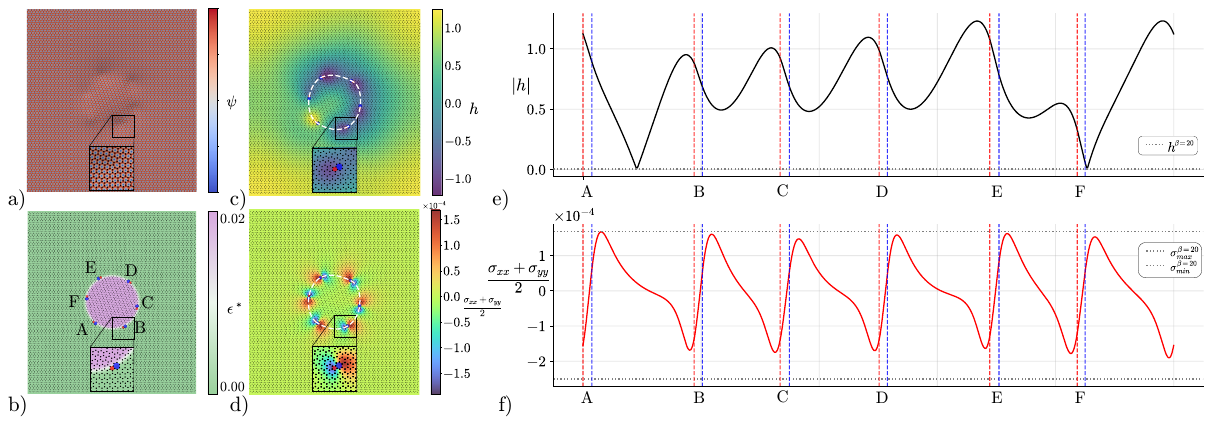}
    \caption{Circular rotated grain with eigenstrain. a) Periodic field $\psi$ on $\mathcal{S}$ after relaxation of the initial condition, allowing for defect formation and out-of-plane buckling. b) Eigenstrain $\epsilon^*$ displayed on $\mathcal{S}$. Dislocations are illustrated as in figure \ref{fig:dislocation}. c) Height profile $h(x,y)$, together with data from b) projected on $(x,y)$-plane. d) Hydrostatic stress $(\sigma_{xx} + \sigma_{yy}) / 2$ together with data from b) projected on $(x,y)$-plane. Insets in a) – d) show zoomed-in views of the region surrounding a single dislocation. The estimated circular interface between the rotated grain and unrotated matrix is shown as white lines in panels c) and d). e) Absolute value of height profile $|h|$ along the interface (white line) in c), with red/blue dashed lines marking the position of the dislocations. f) Hydrostatic stress $(\sigma_{xx} + \sigma_{yy}) / 2$ along the interface (white line) in d), with red/blue dashed lines marking the position of the dislocations. Simulations parameters: $\beta = 0.5$, $R = 10a$ and $\epsilon^* = 0.02$. Other parameters are as in Figure \ref{fig:Eshelby}. The grain is placed in the center of the domain.}
    \label{fig:full1}
\end{figure}

\begin{figure}[h]
    \centering
    \includegraphics[width=1\linewidth]{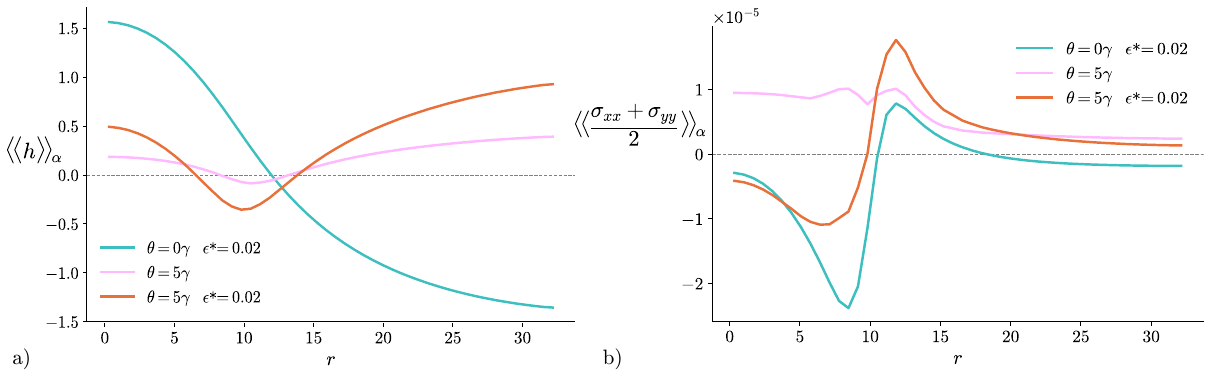}
    \caption{Comparison between different cases. a) Radial average of height profile $\langle \! \langle h \rangle \! \rangle_\alpha$ for the three cases considered in Figures \ref{fig:FvK}, \ref{fig:dislocation} and \ref{fig:full1}, corresponding to ($\theta = 0^\circ$, $\epsilon^* = 0.02$), ($\theta = 5^\circ$, $\epsilon^* = 0$)  and ($\theta = 5^\circ$, $\epsilon^* = 0.02$), respectively. b) Radial average of hydrostatic stress $\langle \!\langle \frac{\sigma_{xx}+\sigma_{yy}}{2}\rangle\!\rangle_\alpha$ for the same cases as in a). $r$ is measured from the center of the domain.}
    \label{fig:full2}
\end{figure}

As the last example, we combine both effects and consider a rotated inclusion with eigenstrain. This setting showcases the general situation of a thin crystalline heterostructure, featuring both lattice misfit and dislocations while deforming out of plane. To be consistent with the previous cases, we consider the same parameters. In Figure \ref{fig:full1}, the corresponding plots to Figure \ref{fig:dislocation} are shown. The stress release in the inclusion superposes here to the bulges at compressive stresses at the dislocations with non-trivial effects on the height profile. While qualitatively similar to Figure \ref{fig:dislocation}, Figure \ref{fig:full1} e) and f) indicate a less regular height profile and hydrostatic stress profile, respectively. 
To compare this setting with the configurations in Figure \ref{fig:FvK} and Figure \ref{fig:Eshelby} we consider the radially averaged height profile and hydrostatic stress. The three cases are shown in Figure \ref{fig:full2}. 
Interestingly, in the case shown here, the configuration combining eigenstrain and dislocations produces a similar buckling pattern, although with a reduced buckling amplitude compared to the defect-free case. This means that, on average, smaller compressive stresses are obtained in the inclusion, while larger tensile stresses are obtained at the interface between the inclusion and the surrounding matrix. We can conclude that, for the considered parameters and initial conditions, defects provide an additional source of strain release and influence out-of-plane deformations. More generally, the approach considered here can be applied to study the interplay between misfit and dislocation in deformable crystalline sheets.

\section{Conclusion}

Lattice mismatch in heterostructures can be modeled as spatially varying eigenstrains. Within the derived sPFC-Helfrich model, changes in the preferred lattice spacing enter through a modulation of the local wavelength of the density field. The model captures the resulting mechanics through a coupling between in-plane distortions of the crystalline lattice and out-of-plane deformations of the surface profile. Both defects and localized variations in the lattice parameter induce out-of-plane deformations, with bulges forming at dislocations where the stress is compressive. This interplay was shown here for a case in which dislocations provide strain release and reduce the amplitude of the buckled surface. 

The model and simulation approach can be applied to larger systems and more complex structures, e.g., polycrystalline metal foils \cite{gautier2025quantifying}. Furthermore, dynamic processes featuring time-dependent eigenstrain and dislocation dynamics over large scales can also be considered. However, reproducing height modulation in the presence of large-angle grain boundaries and under dynamical conditions requires further assessment and investigation, which will be the subject of future work building on the findings reported here.

\begin{acknowledgement}
  E.R. and A.V. acknowledge support from DurAMat under the European Union - Marie Sk\l{}odowska-Curie Actions Doctoral Networks (ITN) Call H2020-MSCA-DN-2022, grant no. 101119767. M.S. acknowledges support from the Deutsche Forschungsgemeinschaft (DFG, German Research Foundation), project numbers 417223351 (FOR3013). The authors gratefully acknowledge the computing time made available to them on the high-performance computer at the NHR Center of TU Dresden. This center is jointly supported by the Federal Ministry of Research, Technology and Space of Germany and the state governments participating in the \href{www.nhr-verein.de/unsere-partner}{NHR}.
\end{acknowledgement}

\appendix

\section{Functional derivatives and notation for height observer } \label{app1}

The functional derivatives in \eqref{eq_heigt1} and \eqref{eq_height2} read:
\begin{align}
\frac{\delta \mathcal{F}_{\mathcal{L}}}{\delta \psi} &=  A  \left( q^4 \psi  + q^2 \Delta_S \psi  + \Delta_S (q^2 \psi) + \Delta_S^2 \psi \right) \\
\frac{\delta \mathcal{F}_{\mathcal{P}}}{\delta \psi} &= B\psi + C \psi^2 + D\psi^3 \\
\frac{\delta \mathcal{F}_b}{\delta \psi} &= 0
\end{align}
and 
\begin{align}
\frac{\delta \mathcal{F}_{\mathcal{L}}}{\delta \xi}&=  \mathcal{H}\frac{A}{2}((\Delta_S \psi)^2 - q^4\psi^2) - 2 A \left(\nabla_S \psi \Pi_{QS} \textit{II} \nabla_S (\Delta_S \psi + q^2\psi) \right)  + \bigg\langle\!\bigg\langle \frac{\delta \mathcal{F}_{\mathcal{L}}} {\delta \psi}\bigg\rangle\!\bigg\rangle  \mathcal{H}\psi\\ 
\frac{\delta \mathcal{F}_{\mathcal{P}}}{\delta \xi} &=  \mathcal{H} \left( \frac{B}{2} \psi^2+ \frac{2C}{3} \psi^3+\frac{3D}{4} \psi^4  \right) \\
\frac{\delta \mathcal{F}_b}{\delta \xi} &= \beta \left( 
\Delta_S \mathcal{H} + \mathcal{H} \left( \frac{\mathcal{H}^2}{2} - 2K \right) \right)
\end{align}
respectively. $\langle \langle.\rangle\rangle$ represents the introduced coarse-graining of the elastic chemical potential over the unit cell. In order to formulate the equations in local coordinates $(x,y)\in\mathcal{U}\subseteq\R^2$ use: $|\mathbf{g}| = 1 + (\partial_x h)^2 + (\partial_y h)^2$, 
$$
\mathcal{H} = \frac{1}{\sqrt{|\mathbf{g}|}}\left( \Delta h 
- \frac{\langle \nabla h, (\nabla^2 h)\,\nabla h \rangle}{|\mathbf{g}|} \right), \quad K = \frac{\det(\nabla^2 h)}{|\mathbf{g}|^2}, \quad \langle \nabla_S f, \nabla_S g \rangle_S = \langle \nabla f , \nabla g \rangle -
\frac{\langle \nabla h, \nabla f \rangle \langle \nabla h, \nabla g \rangle}{|\mathbf{g}|} ,
$$
$$
\nabla_S f = \nabla f 
- \frac{\langle \nabla h, \nabla f \rangle}{|\mathbf{g}|} \nabla h, \quad  \Delta_S f = \Delta f 
- \frac{\langle \nabla h, \nabla f \rangle}{|\mathbf{g}|} \Delta h - \frac{\langle (\nabla h), (\nabla^2 f)\,\nabla h \rangle}{|\mathbf{g}|} + \frac{\langle \nabla h, (\nabla^2 h)\,\nabla h \rangle \langle \nabla h, \nabla f \rangle}{|\mathbf{g}|^2},
$$
$$
\quad \textit{II} = \frac{\nabla^2 h}{\sqrt{|\mathbf{g}|}}, \quad \text{and} \quad \Pi_{QS}\textit{II} = \textit{II} - \frac{\mathcal{H}}{2}\,\mathrm{Id}_{TS}
$$
for $f, g : \mathcal{U} \to \R$ and $\mathrm{Id}_{TS}$ the identity operator in the tangential space.
\vspace{\baselineskip}

\bibliographystyle{pamm}
\bibliography{pamm-tpl}

\end{document}

%% file: symbols.tex
\let\oldcdot\cdot 
\renewcommand{\cdot}{\mathbin{\mkern-2mu\oldcdot\mkern-2mu}}

\newcommand{\eg}{e.\,g.}

\newcommand{\formComma}{\,\text{,}}

\newcommand{\R}{\mathbb{R}}

\newcommand{\surf}{\mathcal{S}}

\newcommand{\tangent}[2][]{\tensor{\operatorname{T}\!}{#1}#2}

\newcommand{\tangentR}[1][]{\tangent[#1]{\R^3\vert_{\surf}}}

\newcommand{\Comp}{\mathsf{C}}

\newcommand{\Div}{\operatorname{Div}}

\newcommand{\DivC}{\Div_{\!\Comp}}

\newcommand{\paraC}{X}
\newcommand{\para}{\boldsymbol{\paraC}}
\newcommand{\normalC}{\nu}
\newcommand{\normal}{\boldsymbol{\normalC}}

\newcommand{\ofrak}{\mathfrak{o}}

\newcommand{\vb}{\boldsymbol{v}}

\newcommand{\Wb}{\boldsymbol{W}}

\newcommand{\vnor}{v_{\bot}}

\newcommand{\DX}{D_{\!\para}}